\documentclass{pasj00}
\draft

\begin{document}
\Received{2010/11/4}
\Accepted{2011/1/24}
\title{Measurements of Antenna Surface for Millimeter-Wave Space Radio Telescope}
\SetRunningHead{K\sc{amegai} \rm et al. }{Antenna Surface}
\author{Kazuhisa \textsc{Kamegai}, Masato \textsc{Tsuboi}, Akihiro \textsc{Doi}, \and Eiichi \textsc{Sato}}
\affil{Institute of Space and Astronautical Science (ISAS), Japan Aerospace Exploration Agency (JAXA), 3-1-1 Yoshinodai, Chuo-ku, Sagamihara, Kanagawa 252-5210, Japan
}
\email{kamegai@vsop.isas.jaxa.jp}

\maketitle

\begin{abstract}
In the construction of a space radio telescope, it is essential to use materials with a low noise factor and high mechanical robustness for the antenna surface. We present the results of measurements of the reflection performance of two candidates for antenna surface materials for use in a radio telescope installed in a new millimeter-wave astronomical satellite, ASTRO-G. To estimate the amount of degradation caused by fluctuations in the thermal environment in the projected orbit of the satellite, a thermal cycle test was carried out for two candidates, namely, copper foil carbon fiber reinforced plastic (CFRP) and aluminum-coated CFRP. At certain points during the thermal cycle test, the reflection loss of the surfaces was measured precisely by using a radiometer in the 41-45 GHz band. In both candidates, cracks appeared on the surface after the thermal cycle test, where the number density of the cracks increased as the thermal cycle progressed. The reflection loss also increased in proportion to the number density of the cracks. Nevertheless, the loss of the copper foil surface met the requirements of ASTRO-G at the end of the equivalent life, whereas that of the aluminum-coated surface exceeded the maximal value in the requirement even before the end of the cycle. 
\end{abstract}
 
\section{Introduction} 
Space Very Long Baseline Interferometry (VLBI) is a form of radio interferometry in which both space-based and ground-based radio telescopes are connected. The first astronomical observations performed by using a space VLBI system were conducted as part of the VLBI Space Observatory Programme (VSOP, \cite{Hirabayashi1998}) with the HALCA satellite (\cite{Hirabayashi2000}). Currently, ISAS/JAXA is developing ASTRO-G, which is a novel satellite featuring a millimeter-wave (mm-wave) radio telescope for use as the VSOP2 interferometer (\cite{Tsuboi2008}, \cite{Saito2008}). The VSOP2 interferometer is designed to achieve an angular resolution of about 40 microarcseconds at 43 GHz, which is 3 times higher compared to any ground-based VLBI system. When deployed, VSOP2 is expected to become the most powerful tool for observing the innermost regions of active galactic nuclei (AGNs) and astronomical masers. 

The antenna of the radio telescope of ASTRO-G consists of a 9.2-m offset-Cassegrain main reflector (Large Deployable Reflector, LDR; \cite{Higuchi2008}), and a 1-m subreflector (SubReflector Unit, SRUNT). The observation frequency bands are X(8.0-8.8 GHz), K(20.6-22.6 GHz) and Q(41-45 GHz). The LDR is designed with a grib hooph back structure and a metal mesh surface, and the SRUNT is designed with a solid surface of carbon fiber reinforced plastic (CFRP). CFRP has been used as a material for covering the surface of onboard communication antennas because carbon fibers are good conductors and fabrics woven from such fibers are good reflectors in the microwave band. In this case, the texture pattern is smaller than the wavelength (\cite{Kaiser1994}). However, the RF performance of CFRP itself does not meet the strict requirements for low noise for radio telescope antennas, especially in the mm-wave band (\cite{Kaiser1994}). Our measurements in the Q-band showed that a bare CFRP surface with no environmental load exhibits a reflection loss as high as $0.22$ dB at 43 GHz, which is significantly higher than the maximal value in the requirement for the SRUNT of ASTRO-G. In this case, the reflection coefficient of the processed surface should be higher. Naturally, the surface material must survive the thermal cycle to which the satellite will be subjected in orbit. In this paper, we present measurements of the reflection loss of the abovementioned candidate materials for the antenna surface of SRUNT together with the parameters of their degradation during the thermal cycle test of ASTRO-G. The reflection loss at the mm-wave band of LDR will be published in a separate paper due to the use of a different measurement technique.

\section{Requirements for Reflection Loss of Space Radio Telescopes}
   There are two applications for an antenna on a satellite, the first of which is communication between the satellite and ground tracking stations and the second is use in radio telescopes, as mentioned in the previous section. 
   
  For both of these applications, the performance of an antenna is evaluated on the basis of $G/T_{sys}$, where $G$ is the antenna gain and $T_{sys}$ is the system noise temperature. The degradation of the reflection coefficient of the antenna surface simultaneously causes the decrease of the antenna gain and the increase in the system noise temperature. Although the antenna gain for these applications is calculated with the same formula, the system noise temperature is evaluated with a different formula depending on the use of the antenna. In the case of a communication antenna, the system noise temperature $T_{sys, c}$ is given by 
           $$T_{sys, c} = T_{earth} + T_{ant}(10^{L/10}-1) + 10^{L/10}T_{rx},$$
where $T_{earth}$ is the brightness temperature of the surface of the Earth, $T_{ant}$ is the physical temperature of the antenna surface, $T_{rx}$ is the noise temperature of the receiver system (e.g., \cite{Otoshi2008}, \cite{Baars2007}), and $L$ dB is the loss depending on the reflection coefficient of the antenna surface. In this case, the $10^{L/10}-1$ in the second term on the right is the product of $10^{L/10}$ and $1-10^{-L/10}$, while the latter is the emissivity of the antenna surface. When both $T_{earth}$ and $T_{ant}$ are assumed to be 300 K, the formula is conveniently expressed as follows:
             $$T_{sys, c} = 10^{L/10} (T_{rx}+300).$$
On the other hand, the noise temperature $T_{sys, r}$ of the radio telescope is given by
            $$T_{sys, r} = 2.7 + T_{ant}(10^{L/10}-1) + 10^{L/10}T_{rx},$$
where the constant 2.7 represents the contribution of the cosmic microwave background.  

   If the loss and the noise temperature of the receiver system are $L=1$ dB and $T_{rx}=50$ K, respectively, $G/T\rm s$ for a communication antenna and a radio telescope degrade to 63\% and 37\%, respectively. The requirement for $G/T$ for the antenna surface of a radio telescope is much stricter than that for a communication antenna. The requirement for the SRUNT of the ASTRO-G satellite is a loss of $0.07$ dB throughout the expected mission lifetime, which corresponds to an additional noise temperature of $3$ K. Therefore, an antenna surface material with a low noise factor and high mechanical robustness is essential for the construction of a space radio telescope.

\begin{figure}
\begin{center}
\includegraphics[width=80mm]{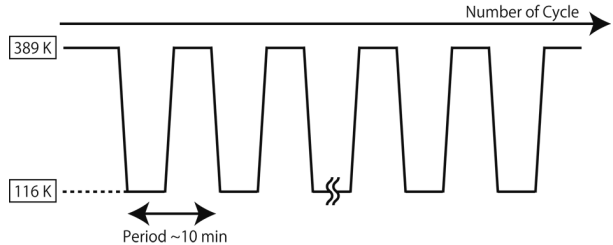}
\caption{Time profile of the temperature in the thermal cycle test. The temperature was controlled in such a way as to confine the thermal cycle within the range between $116 $ K and $389$ K for a total number of 2540 cycles, which corresponds to the equivalent lifetime of the ASTRO-G mission of three years. The test was carried out in a testing facility at ISAS.}
\label{Fig1}
\end{center}
\end{figure}
\begin{figure}
\begin{center}
\includegraphics[width=80mm]{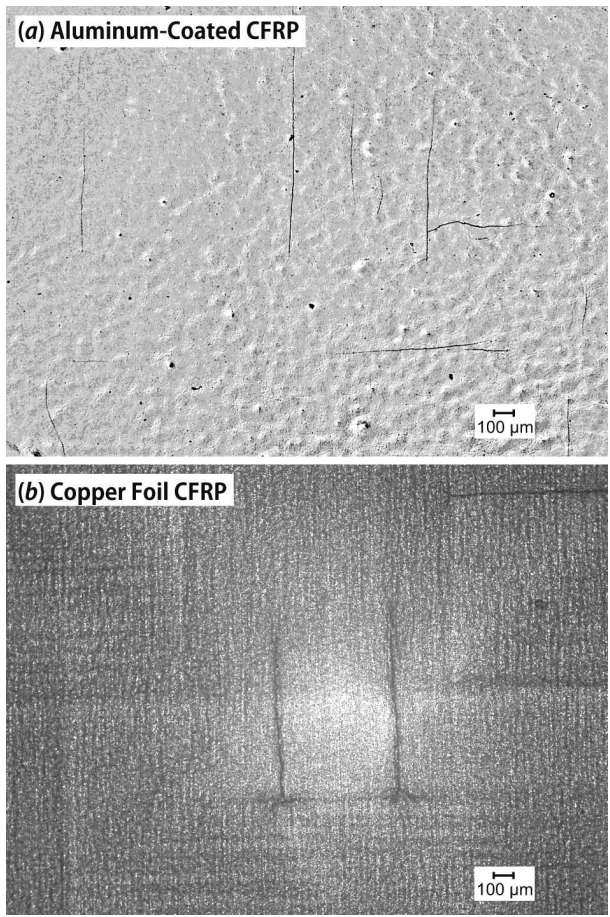}
\caption{ (a) Appearance of the aluminum-coated CFRP surface after 2070 thermal cycles. (b) Appearance of the copper foil CFRP surface after 2070 thermal cycles.}
\label{Fig2}
\end{center}
\end{figure}

\begin{figure}
\begin{center}
\includegraphics[width=80mm]{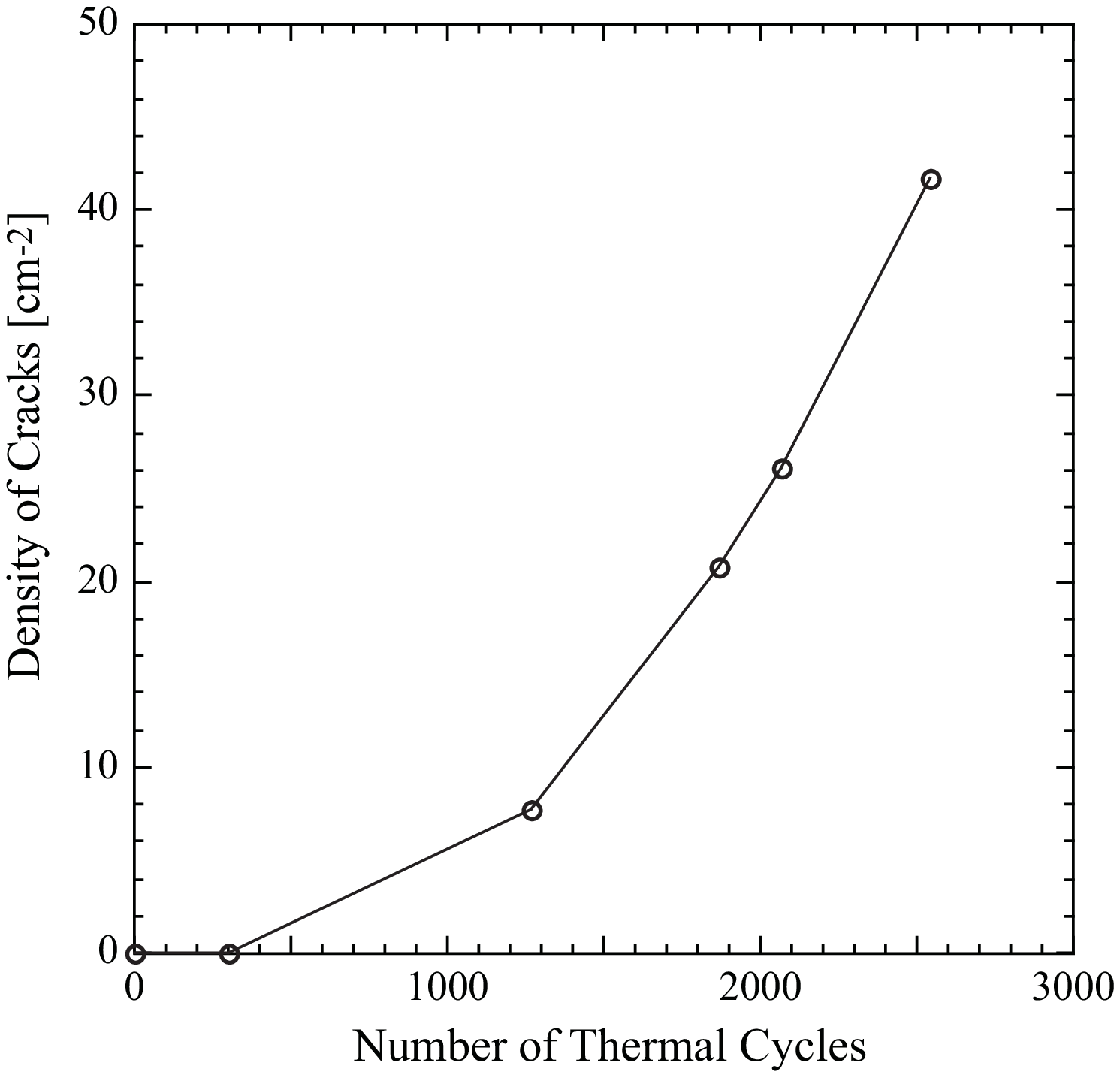}
\caption{Number density of cracks on the copper foil surface as measured in the thermal cycle test. The horizontal axis represents the number of the thermal cycle and the vertical axis represents the density of cracks on the surface. The open circles indicate the results of the measurement.}
\label{Fig3}
\end{center}
\end{figure}

\section{Thermal Cycle Test} 
\subsection{Test Conditions}
  We selected two types of surfaces featuring different candidate materials for the antenna of SRUNT on the basis of existing experience with their fabrication, handling, and other factors. The first type is an aluminum-coated surface, which is fabricated by evaporation in a vacuum chamber. The second type is a copper foil surface, where copper foil and pre-autoclaved CFRP are molded and thermally processed in an autoclave. The specifications of specimens of these surfaces are summarized in Table 1. Here, the flatness of the surface deviates by less than $20 \mu$m from an ideal flat surface.
  
  A thermal cycle test was carried out by taking into account the expected temperature in the environment of ASTRO-G  in its projected orbit. The test was performed in an atmosphere of nitrogen gas with a pressure of 1 atm by using a test facility at ISAS. The temperature environment of SRUNT in orbit depends on various conditions, such as the exact location of the satellite along its orbit, whether the satellite is in the shadow of the Earth, and so forth. The temperature of SRUNT is expected to decrease rapidly to 136 K when the satellite enters the shadow and to increase rapidly to up to 369 K when it exits the shadow. For this reason, the temperature range for the thermal cycle test was set by using the predicted values of the temperature at the surface of SRUNT in orbit with a margin of $\pm 20$ K, namely, $116$ K $< T < 389$ K.
ASTRO-G is projected to follow a unique orbit with large eccentricity in order to achieve both high angular resolution and high image quality with VSOP2. The thermal conditions of such an orbit are much more severe than those of more conventional orbits used with other scientific satellites.
  The period of the thermal cycle was set to ~10 min, as shown in Figure 1. The total number of thermal cycles was 2540, which corresponds to the lifetime of the satellite of three years. The total duration of the test was 27 days (between May 27 and June 22, 2009). 
  We observed the surfaces after 30, 100, 300, 1000, 1270, 1870, 2070, and 2540 cycles under an optical microscope with magnification of $100 - 500$. We also measured the reflection loss of the surfaces at 43 GHz at several points in time during the thermal cycle test. 

\begin{table}
  \caption{Specifications of Specimens. }\label{tab:first}
  \begin{center}
    \begin{tabular}{cll}
    \hline
    Name& Aluminum-Coated CFRP & Copper Foil CFRP\\
\hline
     Size [mm] & $105 \times 97 \times 0.28$ & $210 \times 145 \times 25.6$ \\
      Surface & aluminum, $0.5 \mu$m thickness & copper foil,  $12 \mu$m thickness \\
      Back tructure & cyanate-based CFRP cross-ply & cyanate-based CFRP cross-ply  \\
      & laminated plate (0.07 mm thickness $\times 4$ plies) & laminated plate (0.07 mm thickness $\times 4$ plies) \\
	  & & with aluminum honeycomb panel \\
 \hline
  \end{tabular}
  \end{center}
\end{table}

\subsection{Results of Thermal Cycle Test}
  The thermal cycles gradually introduced mechanical imperfections onto the surface of the specimens. Tarnish, deterioration and cracks were observed on both surfaces during the thermal cycle test. Among these effects, we took particular notice of cracks since they might affect the RF performance. It was discovered that a number of cracks had appeared upon examining the surfaces after the 1000th cycle. At first, cracks appeared as dark lines on the surface after which they gradually grew to the size of cuts on the surface as the heat cycles progressed. Photographs of typical cracks on the surface are shown in Fig. 2. The number of cracks on the copper foil surface was smaller than that on the aluminum-coated surface. Also, all cracks on the copper foil surface were relatively longer but shallower than those on the aluminum-coated surface. Furthermore, the typical length of the cracks was around 1 mm, which is substantially less than the wavelength of 7 mm at 43 GHz. Most cracks appeared only in certain directions, possibly around the boundary of the texture pattern of CFRP. As the number of cycles increased, the length of the cracks also increased, and some of them branched and coupled. We evaluated the mechanical deterioration of the copper foil surface on the basis of the density of cracks which appeared on the surface. The number density of cracks increased as a function of the number of cycles, as shown in Fig. 3.

\begin{figure}
\begin{center}
\includegraphics[width=80mm]{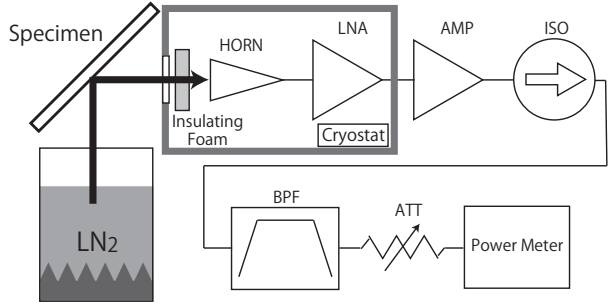}
\caption{Configuration of the apparatus used for conducting the reflection loss measurement. This apparatus consists of two distinct parts, namely, a radiometer and a specimen holder. The radiometer consists of a feed horn, a cooled Low-Noise Amplifier (LNA) placed inside a cryostat, and a power meter operated at room temperature. The LNA is cooled to 30 K with a Gifford--McMahon refrigerator in order to reduce the noise temperature. The frequency range of the measurement is selected with the aid of a waveguide band pass filter (BPF) since the LNA operates at a wide frequency range (between 36 and 50 GHz). The center frequency is 43.0 GHz, and the bandwidth is 4.0 GHz. Also, the load at room temperature is Eccosorb CV-3F, which is located at the input flange of the cryostat. In the case where the load is at the temperature of liquid nitrogen, the feed horn looks at the absorber placed in the Dewar container full of liquid nitrogen through a mirror which is located at the position of the gspecimenh in the figure. We used a flat mirror made of polished aluminum (A5052) and a flat copper foil surface which had not been subjected to thermal cycle tests as mirrors in the respective measurements of the aluminum-coated surface and the copper foil surface. The ATT is a waveguide direct reading variable attenuator, which is used to examine the the linearity of the power meter.}
\label{Fig4}
\end{center}
\end{figure}

\section{Reflection Loss Measurement} 
\subsection{Measurement Configuration}
   We measured the reflection loss of the surfaces at several points in time during the thermal cycle test. As mentioned above, the requirement for low noise for an onboard radio telescope is much stricter than that for a communication antenna. Until this measurement was designed, there had been no measurement apparatus for such low reflection loss at the mm-wave band. Therefore, we developed such a measurement apparatus, which is shown in Fig. 4. This apparatus consists of two distinct parts, namely a radiometer and a specimen holder. The radiometer consists of a feed horn and a cooled LNA placed inside a cryostat, and a power meter operated at room temperature. The LNA is cooled to 30 K with a GM refrigerator in order to reduce the noise temperature. The frequency range of the measurement is selected with a waveguide band pass filter since the LNA covers a wide frequency range between 36 and 50 GHz. In this measurement, the center frequency is 43.0 GHz and the bandwidth is 4.0 GHz, which corresponds to the Q band of ASTRO-G. The feed horn receives the linear polarized signal with horizontal polarization. When the input vacuum flange of the cryostat is alternately terminated with the loads (Eccosorb CV-3F) at room temperature and the temperature of liquid nitrogen, the corresponding outputs from the power meter are measured in order to calibrate the intensity of the injected signal. Here, ATT is a waveguide direct reading variable attenuator, and it is used to examine the linearity of the power meter. The relative accuracy of the calibration is within less than 1\%.
   
  The output $P_{hot}$ terminated with a load at room temperature is given as         
                       $$P_{hot} = G_rk(T_{amb} + T_{rx})B, $$
where $T_{rx}$ and $T_{amb}$ represent noise temperature and room temperature, while $G_r$ and $B$ are the gain of the radiometer receiver and the frequency bandwidth, respectively. On the other hand, the output $P_{cold}$ terminated with a load at the temperature of liquid nitrogen is given as 
                     $$P_{cold} = G_rk(T_{LN2} + T_{rx})B, $$
where $T_{LN2}$ is the temperature of liquid nitrogen.					 
When the loss for the test mirror is $L$ dB, the output from the power meter in brightness temperature is given as        
    $$P_{mirror} =  G_rk[T_{LN2} 10^{-L/10} + T_{amb} (1-10^{-L/10}) + T_{rx}]B. $$ 
The combination of these formulas leads to        
             $$(P_{mirror}- P_{hot})/( P_{cold} -P_{hot}) = 10^{-L/10},$$ 
which yields
$$
             L [dB] =-10\log{\frac{P_{mirror}- P_{hot}}{P_{cold} -P_{hot}}}
$$
We evaluated the loss for the mirror by using the last formula.   
In fact, when we measured $P_{cold}$, the feed horn looked at the absorber located in a Dewar flask full of liquid nitrogen through a mirror with an angle of incidence of $45^{\circ}$. We used a flat mirror made of polished aluminum (A5052) and a flat copper foil surface which had not been subjected to thermal cycle tests as reference mirrors in the respective measurements of the aluminum-coated surface and the copper foil surface. Therefore, the evaluated values of the loss are relative to these mirrors.
\begin{figure}
\begin{center}
\includegraphics[width=80mm]{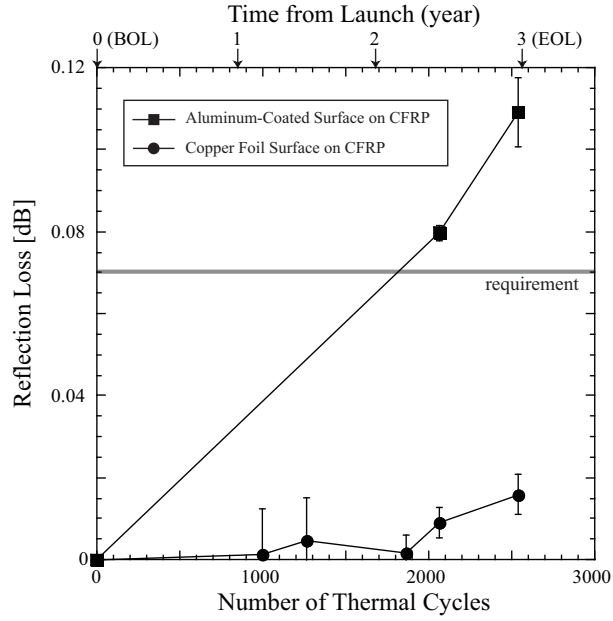}
\caption{Reflection loss during the thermal cycle test. The open circles indicate the copper foil CFRP surface, whose reflection loss gradually increased together with the number of heat cycles. The open squares indicate the aluminum-coated CFRP surface, whose reflection loss increased rapidly together with the number of heat cycles. The gray horizontal line indicates the value corresponding to the requirement for the subreflector unit (SRUNT) of ASTRO-G.}
\label{Fig5}
\end{center}
\end{figure}
\begin{figure}
\begin{center}
\includegraphics[width=80mm]{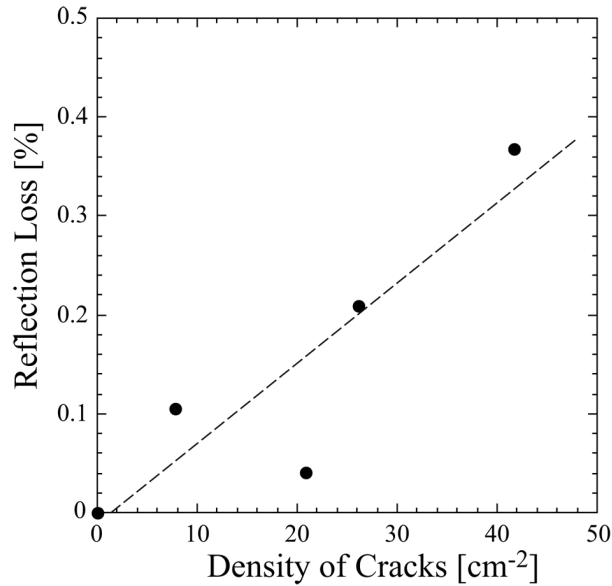}
\caption{Correlation between the reflection loss (in percent) and the number density of cracks on the copper foil CFRP surface. The dashed line indicates the linear function providing the best fit. The reflection loss presumably increases in proportion to the number density of the cracks.}
\label{Fig6}
\end{center}
\end{figure}

\subsection{Reflection Loss Results}
\subsubsection{Copper Foil Surface}
     We measured the reflection loss of the copper foil surface at several points in time during the thermal cycle test, starting with the 1000th cycle. Figure 5 shows the results of these measurements. Although the reflection loss increased gradually as the number of heat cycles increased, it remained as low as $0.016\pm0.005$ dB even at the 2540th cycle, which corresponds to the end of life (EOL) of the ASTRO-G satellite. The error was estimated as the standard deviation of the measurements, which were repeated several times for each cycle. As mentioned above, one of the requirements for the ASTRO-G satellite is a maximal loss of 0.07 dB at the EOL, which corresponds to additional noise temperature of 3 K. The measured loss is considerably lower than the maximal value allowed in the requirement. 
   Both the solar light absorption rate and the infrared radiation factor of the copper foil surface at the EOL were measured to be greater than those at the beginning of life (BOL). After this thermal cycle test, we found that there is a small possibility that the highest physical temperature can be higher than the highest environmental temperature assumed in the test, as well as that the thermal amplitude might increase by 6\%. Based on the theory of metal fatigue, the resultant surface is expected to be equivalent to that at $2540 \times (1.06)^{2} = 2870$ cycles. The result of this simple extrapolation of the reflection loss up to 2870 cycles is expected to remain less than 0.07 dB since the measured loss at 2540 cycles is considerably lower than the maximal value allowed in the requirement (see Fig. 5).
   
   Figure 6 shows the correlation between the reflection loss of the copper foil surface and the density of cracks on the surface. The reflection loss was found to increase in proportion to the number density of cracks. Since CFRP emits in the mm-wave band, some thermal radiation from the CFRP mesh is emitted through the cracks, which presumably acts as a stack of slot antennas (e.g., \cite{Kraus2002}). As the respective emissions from slot antennas are incoherent, an increase in the number density of cracks entails an increase in the number density of such slot antennas, and a sharp synthesis beam is not formed. This leads to an increase in the additional noise at the surface, and is observed as an increase in the reflection loss.
    
Here, a slot antenna is assumed to be a small dipole antenna from Babinet's principle. According to the slot antenna model, a small angle of incidence leads to greater additional noise as it improves the radiation efficiency of the small dipole antenna. In this regard, the radio telescope of ASTRO-G uses circular polarization feeds, where the radiation efficiency of circular polarization is proportional to $1+cos^{2}\theta$. The factor of $1$ here indicates the contribution of small dipole antennas on the surface in orthogonal direction with respect to the beam, while the $ cos^{2}\theta$ factor indicates the contribution of small dipole antennas on the surface in the direction orthogonal to it. On the other hand, the number of cracks in the beam increases as the angle of incidence increases, where the dependence is $1/cos{\theta}$. The additional noise $N_{slot}$ in SRUNT is estimated with this model as
   $$N_{slot}(\theta)=K(1+cos^{2}{\theta})/cos{\theta},$$
where $\theta$ is the angle of incidence, which is assumed to be far from $90^{\circ}$, and $K$ is a constant which depends on the physical temperature and emissivity of the CFRP.

The measurement was performed at an incident angle of $45^{\circ}$ due to a mechanical limitation of the measurement apparatus, as shown in Fig. 4. In the radio telescope of ASTRO-G, the angle of incidence from LDR to SRUNT and from SRUNT to the feed horn is about $36^{\circ}$. In this case, the difference attributable to this additional noise amounts to less than the measurement error.

Therefore, our measurements show that the copper foil surface is a feasible material for millimeter-wave astronomical observations in the thermal environment of ASTRO-G.

\subsubsection{Aluminum-Coated Surface}
    Measurements of the aluminum-coated surface were carried out at cycles 2070 and 2540, where the reflection loss was measured to be $0.079\pm0.002$ dB and $0.109\pm0.008$ dB, respectively (Figure 5). The reflection loss increased rapidly as the number of heat cycles increased. It shows that this type of surface does not meet the RF requirements of ASTRO-G.

\section{Summary}
By employing a thermal cycle test, we performed measurements of the reflection loss of two candidate materials for mm-wave astronomical observations with ASTRO-G. The results can be summarized as follows.
 
(1) The reflection loss of the candidate surfaces deteriorated as the thermal cycle test progressed. Optical microscope observations revealed that cracks on the surfaces increased, became longer and branched.

(2) The reflection loss of the copper foil CFRP surface was measured to be as low as 0.016 dB at the end of the thermal cycle test, which shows that this material is suitable for use as an antenna surface material. On the other hand, the aluminum-coated CFRP surface exhibited a loss as high as 0.11 dB, which is greater than the maximal value in the requirement for ASTRO-G.

(3) The reflection loss of the copper foil surface was found to be correlated with the number density of cracks in the thermal cycle test. It suggests that cracks play the role of slot antennas, and emission of the supporting structure of CFRP has been observed.

\bigskip 
The authors would like to thank Prof. H. Saito, who is the manager of the ASTRO-G project, and Prof. Y. Murata at ISAS/JAXA for their continuous encouragement. The authors also thank Prof. K. Higuchi and Dr. N. Kishimoto at ISAS/JAXA, Prof. H. Ogawa at Osaka Prefecture University, and Prof. T. Kasuga at Hosei University for useful discussion.

\end{document}